\documentclass[aps,pra,floatfix,showpacs,noshowkeys]{revtex4-1}

\usepackage[dvips]{graphicx}
\usepackage{amsmath}
\usepackage{amsfonts}
\usepackage{amssymb}
\usepackage[usenames,dvipsnames]{color}
\usepackage[colorlinks=true,linkcolor=blue]{hyperref}

\newcommand{\newc}{\newcommand}

\newc{\beq}    {\begin{equation}}
\newc{\eeq}    {\end{equation}}
\newc{\beqa}    {\begin{eqnarray}}
\newc{\eeqa}    {\end{eqnarray}}
\newc{\bs}    {\section}
\newc{\no}    {\\ \nonumber}

\newc{\st}    {\stackrel}

\begin{document}
\title{ Holographic Dark Energy and Quantum Entanglement }
\author{Jae-Weon Lee}
\affiliation{ Department of Renewable Energy, Jungwon University,
5 Dongburi, Goesan-eup, Goesan-gun, Chungbuk 367-805, Korea
}

\author{Hyeong-Chan Kim}
\affiliation{School of Liberal Arts and Sciences, Korea National University of Transportation, Chungju 27469, Korea}

\author{Jungjai Lee}
\email{jjlee@daejin.ac.kr}
\affiliation{Division of Mathematics and Physics, Daejin University, Pocheon, Gyeonggi 487-711, Korea}


\begin{abstract}
In this paper, we briefly review the holographic dark energy model and introduce the idea that dark energy is a kind of thermal energy
related to the quantum entanglement of the vacuum across a cosmic future event horizon.
The holographic dark energy model comes from a theoretical attempt to apply the holographic principle to the dark energy problem and follows the idea that the short distance cut-off or ultraviolet (UV) cut-off is related to the long distance cut-off or infrared (IR) cut-off.
The IR cut-off relevant to dark energy is the size of the future event horizon.
This model gives a holographic dark energy comparable to the observational data.
Though this model is in good agreement with observational data, some problems (non-locality, circular logic, causality problem, $\it etc.$) exist due to the use of the future event horizon as a present IR cut-off.
These problems of the holographic dark energy model are considerably resolved using action principle and equations of motion.
Finally, we discuss the relation between quantum entanglement and dark energy which is connected to
the more fundamental relation between entanglement and gravity.
\end{abstract}

\maketitle
\section{Introduction}

The dark energy has become one of the longstanding important central problems in modern cosmology and theoretical physics
ever since the discovery of the accelerating expansion of the Universe in 1998 \cite{riess-1998-116, Perlmutter-1999-517}.
Evidence exists from  Type Ia supernova (SN Ia) observations~\cite{riess-1998-116} that
the  universe is expanding at an accelerating rate, which implies the  existence of dark energy having negative pressure $p_\Lambda$ and density $\rho_\Lambda$ satisfying
the equation of state $\omega_\Lambda\equiv p_\Lambda/\rho_\Lambda<-1/3$.
Many dark energy models, such as the quintessence~\cite{PhysRevLett.80.1582,ZlatevWangSteinhardt:PRL1999,SteinhardtWangZlatev:PRD1999},
$k$-essence~\cite{ChibaOkabeYamaguchi:PRD2000,ArmendarizPiconMukhanovSteinhardt:PRL2000,ArmendarizPiconMukhanovSteinhardt:PRD2001,ArmendarizPiconDamourMukhanov:PLB1999,GarrigaMukhanov:PLB1999},
phantom~\cite{phantom}, Chaplygin gas~\cite{Chaplygin} and
tachyon field~\cite{Padmanabhan:PRD2002,BaglaJassalPadmanabhan2003,AbramoFinelli2003,AguirregabiriaLazkoz2004,GuoZhang2004} models have been suggested; however,
these models usually require a fine-tuning of parameters or unnatural properties of matter to explain the observed data~\cite{CopelandGarousiSamiTsujikawa2005}.
The simplest candidate for dark energy is the cosmological constant with the equation of state $\omega_\Lambda=-1$, which has a theoretical value that is too large to be the observed dark energy~\cite{CC}.

The current observation favor the $\Lambda CDM$ model as the cosmological model. The $\Lambda CDM$ model consists of a cosmological constant $\Lambda$
as the origin of dark energy and a cold dark matter giving rise to galaxies and the large scale structure of the universe.
Although the $\Lambda CDM$ model is favored by observation, it suffers from two cosmological constant problems~\cite{CC,Weinberg1989}: (1) Why is the observed value so small?
(2) Why is its value now on the order of the present matter density?
To solve these problems, numerous models of dark energy have been proposed over the last two decades~\cite{CC,Padmanabhan2003,CopelandSamiTsujikawa2006,FriemanTurnerHuterer2008,CaldwellKamionkowski2009,SilvestriTrodden2009,LiLiWangWang2011,BambaCapozzielloNojiriOdintsov2012}.
However, the nature of dark energy still remains a deep mystery.
We believe that the dark energy problem is essentially an issue of quantum gravity.
Only a full understanding of quantum gravity through a well-established quantum theory can solve the dark energy problem.
The most fundamental principle of quantum gravity is thought to be the holographic principle, which may play an essential role in solving the dark energy problems.

A. Cohen {\it et al.}~\cite{CohenKaplanNelson:PhysRevLett.82.4971} suggested that in the effective quantum field theory, a long-distance infrared (IR) cut-off is related to a short-distance
ultraviolet (UV) cut-off due to the bound of energy in a region of size $L$ by the formation of a black hole of the same size $L$,
which is inspired by the existence of the limited bound for the total entropy $S$ of the system of the same size $L$ in blackhole thermodynamics~\cite{Bekenstein1973,Hawking1975}.
These interrelations between the UV cut-off and the IR cut-off are called UV/IR mixing,
which may be demonstrated explicitly in noncommutative field theory~\cite{MinwallaRaamsdonkSeiberg1999} and, in particular, in emergent quantum gravity~\cite{HSYang,LeeYang2014}.
In 2004, by applying the holographic principle to dark energy, based on the idea of Cohen {\it et al.}~\cite{CohenKaplanNelson:PhysRevLett.82.4971}, Li~\cite{Li2004-603}
suggested the holographic dark energy model in which the IR cut-off is chosen as the size of the future event horizon of the Universe.
The holographic dark energy can drive the accelerating expansion of the Universe and is in good agreement with the present cosmological observational data~\cite{LiLiWangZhang2009}.
Thus, the holographic dark energy has now become one of the most competitive candidates for dark energy~\cite{WangWangLi:2017}.

However, the above approach brings about some criticisms due to the use of the future event horizon as the present IR cut-off. These criticisms are summarized as the causality problem, circular logic problem and non-locality problem.
The causality problem occurs due to the fact that the evolution of the Universe depends on the future information for the Universe because the future even horizon is defined on the future information of the Universe.
The circular logic problem states that the future event horizon exists only in accelerating universe.
In addition, the equations of motion become non-local because the future event horizon is defined globally.
Fortunately, these problems are resolved with the action principle ~\cite{LiMiao:2012} and the equation of motion including the initial value~\cite{KimLeeLee2013}.
In Ref.~\citealp{LiMiao:2012}, by introducing two new fields in the action, holographic dark energy is shown to obey the causality
because the future event horizon as a present cut-off is not an input but is automatically determined by the present cut-off from the equations of motion.
However, there unsatisfactory points still remain because the action is not written in a general covariant form. Complete resolutions of these problems might be expected for a complete quantum gravity theory.

In the recent decade, the holographic paradigm of dark energy has been widely studied, and many theoretical models including action principle~\cite{LiMiao:2012},
such as entanglement entropy~\cite{MullerLousto1995}, holographic gas~\cite{LiLiLinWang2009}, Casimir energy~\cite{CasimirPolder1948,FischettiHartleHu1979,HartleHu1979,HartleHu1980,LiMiaoPang2010-55,LiMiaoPang2010-9026},
and entropic force~\cite{Verlinde2011,LiWang2010}, have been proposed to theoretically explain the origin of holographic dark energy. In particular, that links exist between gravity and quantum information (quantum entanglement, especially)~\cite{LeeLeeKim2015}
or, more specifically, between dark energy and the quantum entanglement of the vacuum of the Universe is very interesting~\cite{LeeLeeKim2007}.

Quantum entanglement~\cite{nielsen} is a physical resource for  the quantum key distribution and quantum computing and is an order parameter for certain condensed matter~\cite{PhysRevA.66.032110}.
Recently, interest in quantum entanglement in the theoretical physics community has been growing~\cite{VanRaamsdonk:2016exw,Brustein:2000,Takayanagi:2012}.
For example, entanglement was proposed to be the source of gravity \cite{Lee:2010bg,Faulkner:2013ica,Oh:2017pkr} and spacetime~\cite{VanRaamsdonk:2010pw}.
Before these works, studies on the role of entanglement in cosmology had been published. (See  Refs.~\cite{MartinMartinez:2012sg,MartinMartinez:2014sg} for a review.)
Those works usually focused on entanglement generation during inflation ~\cite{Nambu:2008my}.

In this paper, we review briefly the dynamics of dark energy and describe the properties of the holographic dark energy model.
In particular, we explain the idea linking holographic dark energy with vacuum entanglement~\cite{LeeLeeKim2007}.
The holographic dark energy model is based on the holographic principle proposed by 't Hooft and Susskind~\cite{hooft-1993,Susskind:1995} and may solve the cosmological constant problem.
A fundamental connection between gravity and quantum information (quantum entanglement, especially) has been proposed in a series of works~\cite{LeeLeeKim2007,Kim:2007vx,Kim:2008re,Lee:2010bg,kias}.
For example, the Einstein equation for gravity was suggested to be equal to the entanglement first law
by considering quantum entanglement entropy of local Rindler horizons based on Jacobson's idea linking the Einstein equation to the first law of thermodynamics~\cite{Lee:2010bg,kias}.

In Section II, we briefly introduce the basic knowledge of modern cosmology, such as  the Friedmann-Lemaitre-Robertson-Walker (FLRW) cosmology, and the dynamics of dark energy.
Section III is devoted to a discussion of the properties of and the problems in holographic dark energy.
In Section IV, we analyze the causality problems in holographic dark energy in detail.
In Section V, we describe the idea linking quantum entanglement to holographic dark energy. Section V contains conclusions and discussions.
Throughout the review, we use natural units $c=\hbar = 1$ and a metric signature $(-,+,+,+)$.

\section{Modern Cosmology and Dark Energy}

In this section, we will give a brief introduction to the theoretical base of modern cosmology and dark energy.
\subsection{FLRW cosmology}
Modern cosmology is based on two fundamentals. The first is Einstein's equations which describe the dynamics of the universe:
\beq \label{EinsteinEquation}
G_{\mu \nu} = R_{\mu \nu} - \frac{1}{2} g_{\mu \nu} R = 8 \pi G T_{\mu \nu},
\eeq
where $G_{\mu\nu}$, $R_{\mu\nu}$, $g_{\mu\nu}$ and $R$ represent the Einstein tensor, the Ricci tensor, the metric tensor, and the Ricci scalar, respectively, $T_{\mu\nu}$ is the energy-momentum tensor, and $G$ denotes the universal gravitational constant.
Let us consider an ideal perfect fluid as the source for the energy-momentum tensor $T_{\mu\nu}$. In this case, the energy-momentum tensor is given by
\beq \label{EnergyMomentum}
T_{\mu \nu} = (\rho + p)u_{\mu} u_{\nu} + g_{\mu \nu} p.
\eeq
Here, $\rho$ is the total energy density, $p$ is the pressure density of the fluid, and $u_{\mu}$ is the fluid four velocity.

The second is the cosmological principle, which is based on the assumption of isotropy and homogeneity of the Universe, which is true approximately on large scales.
Based on the presence of generic symmetries, the Universe can be described by using FLRW metric~\cite{LiddleLyth2000, Weinberg1972,KolbTurner1990,Padmanabhan2000}:
\beq \label{FLRWmetric}
ds^2 = - dt^2 + a^2 (t) \left[ \frac{dr^2}{1-Kr^2} + r^2 (d\theta^2 + \sin^2 \theta d\phi^2 )\right].
\eeq
Here, $a(t)$ is the scale factor with the cosmic time $t$, and the coordinates $r$, $\theta$ and $\phi$ are the comoving coordinates. A freely moving body comes to rest under these coordinates.
The constant $K$ characterizes the geometry of 3-dimensional space, where closed, flat and open universe correspond to $K=+1, 0, -1$, respectively.

By inserting Eq. (\ref{FLRWmetric}) into Eq. (\ref{EinsteinEquation}), we can get the two Friedmann equations:

\beq \label{FriedmannEq1}
3 M^2_p H^2 =\rho -\frac{3 M^2_p K}{a^2},
\eeq
\beq \label{FriedmannEq2}
\frac{\ddot{a}}{a} = - \frac{1}{6 M^2_p} (\rho + 3 p).
\eeq
Here, we have defined the Hubble parameter $H\equiv \dot{a}/a $, which determines the expansion rate of the universe, and the dot represents the derivative with respect to the cosmic time t.
For the late analysis, instead of $G$, the reduced Planck mass $M_p = \sqrt{1/8 \pi G}$ is used .
From the two Friedmann equations, Eqs. (\ref{FriedmannEq1}) and (\ref{FriedmannEq2}), we can get the continuity equation, which describes the conservation of the energy-momentum tensor:
\beq \label{continuityEq}
\dot{\rho} + 3 H ( \rho + p)=0.
\eeq

The expansion rate of the universe may be, in principle, positive or negative; these correspond to expansion or contraction of the universe, respectively.
In 1929, Hubble's astronomical observation showed that the universe is expanding~\cite{Hubble:1929}, and in 1998, Riess {\it et al.} and Perlmutter {\it et al.} discovered that the universe is expanding at an increasing rate~\cite{riess-1998-116, Perlmutter-1999-517}.
In a theoretical aspect, from the second Friedmann equation, we can see that the pressure $p$ will affect the cosmic acceleration expansion. With the definition of the equation of state $w$ in $p = w \rho$, if $w < - 1/3$, the universe will accelerate.

\subsection{Dark energy}

The energy density on the right-hand side of the two Friedmann equations includes all components present in the Universe, namely, radiation, baryon matter, dark matter, and dark energy:
\beq
\label{energyDensity}
\rho = \sum_i \rho^{0}_i (1+z)^{3(1+w_i)},
\eeq
where $\rho^{(0)}_i$ and $w_i$ correspond to the present energy density and the equation of state of each component, respectively.
By introducing the dimensionless density parameter $\Omega(t) = \rho (t) / \rho_c (t) $ with the critical density $\rho_c (t)=3 M^2_p H^2 (t)$, We can rewrite Eq. (\ref{FriedmannEq1}) as
\beq
\label{EqDensityParameter1}
H^2 = H^2_0\sum_i \Omega^{0}_i (1+z)^{3(1+w_i)},
\eeq
where $H_0$ and $\Omega^{0}_i =\rho^0_i / \rho^0_c $ are the Hubble parameter and the density parameter for each component at the present epoch, respectively.
Here, $z=-1+ a_0 / a = -1 + a^{-1}$ is the red-shift parameter with the present scale factor $a_0$ taken to be $ 1$.
Obviously, $\sum_i \Omega^0_i = 1$.
The explicit form of Eq. (\ref{EqDensityParameter1}) for each component of the universe is
\beq
\label{EqDensityParameter2}
H = H_0 \sqrt {\Omega^{0}_r (1+z)^4 + \Omega^{0}_b (1+z)^3 + \Omega^{0}_{DM} (1+z)^3 + \Omega^{0}_K (1+z)^2 + \Omega^{0}_{DE} X(z)},
\eeq
where $\Omega^0_r$, $\Omega^0_b$, $\Omega^0_{DM}$, $\Omega^0_K$ and $\Omega^0_{DE}$, denote the present density parameter for each component: radiation, baryon matter, dark matter, spatial curvature and dark energy, respectively.
Here the dark energy density function is defined from the continuity equation, Eq. (\ref{continuityEq}), as
\beq
\label{DEfunction}
X(z) \equiv \frac{\rho_{DE} (z)}{\rho^0_{DE}}= exp[3 \int^z_0 dz \frac{1+ w(z)}{1+z}],
\eeq
where $w \equiv p_{DE} / \rho_{DE}$ is the equation of state of dark energy.
With the total fraction of matter density $\Omega_M = \Omega_b + \Omega_{DM}$ and the effective energy density of spatial curvature $\rho_K = - 3 M^2_p K / a^2$,
the most important quantity characterizing the properties of dark energy is the dark energy function $X(z)$~\cite{HutererStarkman2003, HutererCooray2005,HuangLiLiWang2009,WangLiLi2011,LiLiWangZhangHuangLi2011,WangHuLiLi2016}.

In fitting the current observational data so far, the $\Lambda CDM$ model, which is the simplest model with the equation of state $w=-1$, still gives the best performance,
so this model is viewed as the standard model of physical cosmology.
However, the standard model of cosmology has two cosmological constant problems~\cite{Padmanabhan2002,Padmanabhan2005}: (1) Why is $\rho_\Lambda \approx 10^{-120} M^4_p$ so tiny?
This poses the most severe naturalness problem in theoretical physics, which is the disagreement between the theoretical large value of the zero-point energy suggested by quantum field theory and the observed small values of the vacuum energy density.
(2) Why now is $\rho_\Lambda \approx \rho_M$? While a cosmological constant is by definition time-independent, the matter energy density is diluted as $a^{-3}$ as the Universe expands.
This is called a coincidence problem. Despite many challenges of the past two decades to solve the problem of dark energy, unfortunately, the nature of dark energy still remains unresolved.

In fact, because the dark energy problem is essentially a matter of quantum gravity, only a full understanding of quantum gravity can solve this problem.
Therefore, the holographic principle as the most fundamental principle of quantum gravity will surely play an important role in solving the dark energy problem.
The holographic dark energy model based on the holograpic principle is likely to present a fundamental answer to this problem because this model is in good agreement with the current observational data.

\section{Holographic dark energy}
Many dark energy models based on quantum field theory (QFT), such as the quintessence~\cite{PhysRevLett.80.1582},
$k$-essence~\cite{ArmendarizPiconMukhanovSteinhardt:PRL2000}, and phantom~~\cite{phantom} models, exist.
However, in these models, the zero-point energy of quantum fields given by
\beq
\rho_\Lambda\sim \int^\Lambda_0 k^2 dk \sqrt{k^2+m^2}\sim {\Lambda^4}
\eeq
contributes to the cosmological constant about $O(10^{70}) GeV^4$  for the UV cutoff $\Lambda\sim M_P$,
where $M_P$ is the Planck mass.
This value is $O(10^{120})$ times larger than the observational data. In short, the number of degrees of freedom of QFT
is too large for dark energy to explain the observational data.
This difficulty can be overcome when we consider the holographic principle by saying that the actual number of degrees of freedom of a region is proportional to its area not to its volume.

From this viewpoint, $\rho_\Lambda$ is a function of the square of the IR length scale $L$ and the  UV cutoff scale $\Lambda$.
From dimensional analysis, we expect
\beq
\rho_\Lambda=\sum_{i=0}C_i M_P^{4-2i}L^{-2i}.
\eeq
The $C_0$ term is the problematic vacuum energy and the terms with $C_i$ $(i\ge 2)$ are negligible
compared to the $C_1$ term.
Hence, dark energy with $\rho_\Lambda=C_1 M_P^{2}L^{-2}$ is a reasonable guess, which
 is further justified by Cohen {\it et al.}~\cite{CohenKaplanNelson:PhysRevLett.82.4971}
who considered the fact that the quantum vacuum energy in a region of size $L$,  $O(L^3\Lambda^4)$, cannot be larger than the black-hole mass of the same size, $O(LM_P^2)$; thus, we have
\beq
 \label{energybound}
L^3\rho_\Lambda \leq LM_P^2.
 \eeq
Saturating this bound, Eq. (\ref{energybound}) again gives  $\rho_\Lambda=C_1 M_P^{2}L^{-2}$.
Therefore, imagining a holographic dark energy model with
 \beq
 \label{holodark}
\rho_\Lambda=\frac{3 d^2 M_P^2}{ L^2 },
 \eeq
 where  $d$ is an $O(1)$  parameter characterizing the  equation of state, is natural.
One can easily derive a useful relation from the Friedmann equation $\rho=3M^2_P H^2$:
\beq
HL=d,
\eeq
which holds for a flat and dark energy dominated universe.

 A natural length scale for cosmology is the Hubble radius $L=H^{-1}$, and
$\rho_\Lambda=3 d^2 M_P^2 H^2$ is a simple guess for holographic dark energy.
Interestingly, the present Hubble parameter  $H=H_0\sim 10^{-33}~eV$ gives a holographic dark energy
comparable to the observed value $\rho\sim
10^{-10} eV^4$. The success of this simple estimate is quite remarkable. Compared to other models, holographic dark energy models do not need an  $ad~hoc$ canceling mechanism
or fine-tuning to solve the cosmological constant problem.
However, Hsu~\cite{hsu} pointed out that the Friedmann equation $\rho=3M^2_P H^2$ for Hubble radius requires holographic dark energy to behave like ordinary
matter and does not give an  expansion of the universe.
Huang and Li~\cite{1475-7516-2004-08-013} showed that if the future event horizon ($R_h$) is used for $L$, holographic dark energy  of the form
\beq \label{holodark2}
\rho_\Lambda=\frac{3 d^2 M_P^2}{ R_h^2 },
\eeq
 can  give an accelerating universe.
 Here,
 \beq
 R_h=a\int^\infty_t \frac{dt'}{a}=a\int^\infty_a \frac{da'}{Ha'^{2}}=\frac{d}{H}.
 \label{Rh0}
 \eeq
 Differentiating $R_h/a$ with respect to $a$ gives the differential equation
 \beq
 -\frac{1}{Ha^2}=d\frac{d}{da}\left(\frac{1}{Ha} \right),
 \eeq
 which has a power-law solution $H=\alpha a^{-1+1/d}$.
 Therefore, $R_h=d/H\propto  a^{1-1/d}$, which yields the dark energy density
 \beq
 \rho_\Lambda=3 \alpha^2 M_P^2  a^{-2(1-1/d) },
 \eeq
which has an equation of state $\omega=-1/3-2/(3d)$.
Thus, for $d>0$, holographic dark energy gives an accelerating expansion of the Universe.
If we include other matter, the  equation of state at the present for
holographic dark energy  becomes  ~\cite{Li2004-603,1475-7516-2004-08-013}
\beq
\label{omega}
w_0 =-\frac{1}{3} \left(1+\frac{2\sqrt{\Omega^0_\Lambda} }{d}\right),
\eeq
and
its  change rate  is ~\cite{Li2004-603,HuangGong2004}
\beq
\label{omega3}
w_1
 =
\frac{\sqrt{\Omega^0_\Lambda} \left( 1 - \Omega^0_\Lambda  \right)}{3d} \left(1+\frac{2\sqrt{\Omega^0_\Lambda} }{d}\right),
\eeq
where  $\Omega_\Lambda^0$ is the density parameter of holographic dark energy at the present.

Observational data for dark energy are usually given with a parametrization like
 $w_\Lambda (z) \simeq w_0+w_1 (1-R)$, where  $R$ is the scale factor of the Universe at the redshift $z$.
For the fiducial value  $\Omega^0_\Lambda= 0.688$ and $d=1$, Eqs. (\ref{omega}) and  (\ref{omega3}) give $w_0=-0.886$ and $w_1=0.229$.
If $d = 1$, $w_\Lambda (z)$ asymptotically approaches $-1$, and the Universe becomes a de Sitter-like universe.
If  $d < 1$, $w_\Lambda (z)$ will cross the  $w_\Lambda (z)=-1$ boundary, and holographic dark energy acts as a phantom dark energy with a big rip.
If  $d > 1$, $w_\Lambda (z)>1$ and holographic dark energy acts as quintessence.
We can determine the value of $d$ by using the current observational data, which favor $d$ smaller than $1$.
For example, a combination of the Planck data, the  baryon acoustic oscillation (BAO), and Type Ia supernovae (SN)
yields a matter density parameter $\Omega_{m0}=0.288^{+0.015}_{-0.013}$ and $d=0.768^{+0.112}_{-0.068}$~\cite{Wang:2013zca}.

Although the holographic dark energy model is in good agreement with current observational data, this model causes fundamental criticisms because it uses the future event horizon as an IR cut-off.
These problems are summarized in three major problems. The first is the causality problem: the evolution of the Universe depends on the future information of the universe.
The second is the circular logic problem: the future event horizon exists only in the accelerating Universe. The third is the non-locality problem: the equation of motion is non-local because the future event horizon is defined globally.
It has been shown that these problems can be resolved by the action principle and an analysis of equations of motion~\cite{LiMiao:2012,KimLeeLee2013}.

\section{Causality of the Holographic Dark Energy Model}
In this section, the attempts to resolve these problems based on holographic dark energy itself rather than introducing other interactions are summarized in some detail.
The distance to the future event horizon in a flat Robertson-Walker spacetime is given by
\begin{equation} \label{Rh}
R_h(t) \equiv a(t) \int_t^\infty \frac{dt'}{a(t')} .
\end{equation}
The equation of motion of the Universe in the presence of holographic dark energy is given by
\begin{equation} \label{HDE}
3M_p^2 H^2 = \rho_\Lambda + \rho_{nh},
\end{equation}
where $\rho_\Lambda$ represents the energy density of the holographic dark energy in Eq.~\eqref{holodark2} and $\rho_{nh}$ represents the energy density of all other forms of matters satisfying local equations of motion.

A major objection to the holographic dark energy model comes from the form of the holographic dark energy in Eq.~\eqref{holodark2} because it depends on the future evolution of the Universe as in Eq.~\eqref{Rh}.
Because of the future dependence, doubt arises that it may violate causality or has a circular logic problem because the future event horizon is defined globally~\cite{Cai2007} as in Eq.~\eqref{Rh}.
The problems can be summarized as follows:
\begin{itemize}
\item {\it Causality problem:} Assume that a creature who can modify the future event horizon exists.
Because the equation of motion depends on the size to the future event horizon, if the horizon is modified, that may affect the present motion of the universe.
This raises the problem of causality: "How can we expect the next state of the universe from our current data when we have no future knowledge?"
\item {\it Circular logic:} Given the current data on the Universe, we cannot know whether the Universe will eventually be in the phase of accelerating expansion or not.
The future event horizon is not determined until the Universe finishes its final evolution.
Once holographic dark energy is introduced, on the other hand, the universe is destined to expand at an accelerating rate.
If we do not know the present existence of the future event horizon, how can we use it?
This constitutes the essence of the circular logic problem. How can we use an assumption (future event horizon) based on the accelerating expansion to explain the accelerating expansion.
When we point out that holographic dark energy based on horizons other than the future event horizon does not give the accelerating expansion of the Universe, the problem becomes worse.
\end{itemize}

Various attempts have been made to overcome these shortcomings.
In the Brans-Dicke theory of gravity, by adopting the Hubble scale as the IR cut-off instead of the future event horizon, Gong {\it et al.}~\cite{gong:064029,GongLiu:2008} developed the extended holographic dark energy model.
A no-go theorem exists that the Hubble scale cannot be chosen as an IR cut-off for a Universe governed by the Brans-Dicke theory of gravity~\cite{gong:064029,GongLiu:2008}.
However, given a potential term for the scalar field in the Brans-Dicke gravity, Liu {\it et al.} succeed to prove that the holographic dark energy can be generated with the Hubble horizon~\cite{Liu:2010}.
Many authors considered different IR cut-off scales, such as the Ricci curvature radius~\cite{Nojiri:2005pu} or the age of the Universe~\cite{Cai2007,Wei2008}.
Other attempts~\cite{horvat:087301,0295-5075-71-5-712,gong:064029,pavon-2005-628} use gravity action~\cite{LiMiao:2012}, non-minimal coupling
~\cite{0295-5075-71-5-712,gong:064029} or the interaction between dark energy and dark matter ~\cite{pavon-2005-628}.

\subsection{Properties of the future event horizon}
Let us briefly summarize the properties of the future event horizon.
In order to understand causality, even though co-moving time is intrinsic to a co-moving observer, co-moving time may not be the best parameter to describe time.
Based on a conformal time measured from infinity,
\begin{equation} \label{eta}
\eta = \int_\infty ^t \frac{dt'}{a(t')},
\end{equation}
the event horizon appears to be causal, as discussed by Li in Ref.~\citealp{Li2004-603}.
Adopting that time, the flat Robertson-Walker metric takes the form
\begin{equation} \label{metric:2}
ds^2 = a^2(\eta) ( - d\eta^2 + d{\bf x}^2) .
\end{equation}
Now, the range of conformal time has a finite upper bound, for instance $\eta \in (-\infty,0)$.
Due to this finite upper limit, a light ray starting from the origin at time $\eta$ cannot reach a certain distance arbitrarily, but has a horizon at $ r = -\eta$.
Now, the formula for the distance of the future event horizon, $R_h = a(\eta)|\eta|$, appears to be causal.
This justification, however, resorts to the existence of a peculiar conformal time, defined by the integral from the future infinity.
Therefore, the circular reasoning problem remains unsolved.

Second, the future event horizon is always located at outside the Hubble horizon if $\dot R_h > 0$, as can be seen from the equation satisfied by the distance to the horizon,
\begin{equation} \label{dRh}
H^{-1}\dot{R}_h = R_h -H^{-1},
\end{equation}
as given in Ref.~\citealp{Kim:2007kw}.
All observers in the universe will be surrounded by future event horizons, and access to remote information in the Universe will be limited to the horizon.
The absence of this information is represented by a kind of entropy given by the horizon area in Planck unit; which is similar to the black hole entropy.
Davies~\cite{Davies:1988} and Pollock and Singh~\cite{Pollock:1989} proved the generalized second law of thermodynamics for the Robertson-Walker space time, which states that the total entropy of gravity and matter never increases through physical process.
In a sense, Eq.~\eqref{dRh} provides a clue to the resolution of the causality problem because this equation implies that the horizon satisfies a causal evolution equation even though horizon itself is defined globally.

\subsection{Resolution of the causality problem based on the equation of motion}

The circular logic problem is rooted in the assumption that the future accelerating expansion is a natural consequence because the holographic dark energy comes from the existence of the future event horizon.
However, the universe may not be destined to expand at a rate that accelerates in the future, as shown in Ref.~\citealp{KimLeeLee2013}, even though the energy density of the Universe depends on the future event horizon.
This implies that the future accelerating expansion is not a natural consequence of the existence of a future event horizon.
Therefore, a breakthrough must exist in the problem. The discussions in this subsection follows the work in Ref.~\citealp{KimLeeLee2013}.

Let us specify the origin of the causality problem in the evolution equation, Eq.~\eqref{HDE}.
Then, we isolate it as a boundary condition from a well-posed differential equation, which can be determined from initial data.
Suppose that we have a creature that can modify the future event horizon, and someday in the future, he  changes the horizon.
Now, we can ask the following two questions:
\begin{enumerate}
\item Does this behavior modify the current evolution of the Universe in which the creature lives?
\item Does this behavior imply a violation of causality for us?
\end{enumerate}
The answer to the first question must be ``yes".
The evolution equation justifies the statement that the modification of the horizon area actually changes the current evolution.
However, this does not make the answer to the second question also ``yes".

To answer the second question, we review the Eq.~\eqref{HDE} carefully.
We separate the future-dependent part from the true dynamics described by a well-posed second-order differential equation.
By inserting the future event horizon into the holographic dark energy, we can rewrite the equation of motion, Eq.~\eqref{HDE}, as
\begin{equation} \label{HDE:2}
\int_t^\infty \frac{dt'}{a(t')}
	= \frac{d}{a(t) \sqrt{H^2 - \rho_{nh}/3M_p^2 }} ,
\end{equation}
where we assume $d> 0$.
Note that the future-dependent part can be localized on the left-hand side of Eq.~\eqref{HDE:2}.
We can separate that part from the others by setting $\int_t^\infty dt'/a(t') = r_\infty - \int_0^t dt'/a(t')$, where $r_\infty = \int_0^\infty dt/a(t)$ bears the future dependence of the equation of motion.
When Eq.~\eqref{HDE:2} is differentiated with respect to $t$, the term $r_\infty$ disappears, and well-posed second-order differential equation is obtained:
\begin{equation} \label{HDE:3}
\dot H = H^2- \frac{\rho_{nh} - \dot{\rho}_{nh}/2H }{3 M_p^2} +
	\frac{[H^2 - \rho_{nh}/(3M_p^2)]^{3/2}}{d H} .
\end{equation}
Finally, we divide the evolution equation, Eq.~\eqref{HDE}, into two pieces; one is an evolution equation~\eqref{HDE:3}
and the other is the term that appears to contain future information, $r_\infty$.

Note that the value $r_\infty$ does not affect the evolution equation, Eq.~\eqref{HDE:3}.
Therefore, all evolution can be resolved from some initial information without any future information from Eq.~\eqref{HDE:3}.
Because the evolution of the Universe can be determined in the absence of $r_\infty$,
solving the equations of motion, Eq.~\eqref{HDE:3}, reveals that the horizon can be determined from current information.
In this sense, we do not need information about $r_\infty$ to determine the distance to the event horizon.
In fact, $r_\infty$ is not always well defined.
It is finite only when the scale factor behaves as $a(t) \sim t^m ~(m< 1)$ for small $a$ and as $a(t) \sim t^n~(n> 1)$ in the future.
However, even if $r_\infty$ is ill-defined, the evolution equation, Eq.~\eqref{HDE:3}, works well and reproduces the Einstein equation.
If someone says that causality has been violated, he means that from the past data, something unexpected happens at present (due to the future action).
However, as seen in Eq.~\eqref{HDE:3}, all future evolutions are predictable from the past data.
If `causality' is defined in this sense, then the answer to the second question is ''No''; 'causality' is not violated for us.

Now, let us discuss the role of $r_\infty$.
Even though $r_\infty$ is defined by the integral over time, we may also obtain its value by using the limit
\begin{equation} \label{rinfty}
r_\infty = \lim_{\epsilon \to 0} \int_\epsilon^\infty \frac{dt'}{a(t')}
	= \lim_{\epsilon \to 0} \frac{d}
		{a(\epsilon)\sqrt{H(\epsilon)^2 - \rho_{nh}(\epsilon)/3M_p^2}} .
\end{equation}
Therefore, rather than integrating over the whole evolution, the limit to obtain $r_\infty$ may simply be taken.
From this point of view, the value $r_\infty$ plays the role of an initial boundary condition rather than a constant dependent on the future.
Summarizing, if a creature modifies the future event horizon, that simply means that he manages to change the initial condition given as $r_\infty$ or the boundary condition.
In that case, the creature may move from one universe to another universe (multi-universe theory).
On the other hand, an observer in a universe may not recognize the change in the initial condition because he/she has been living in the modified universe.
Rather, the observer will see a universe that follows the evolution equation, Eq.~\eqref{HDE:3}, with a modified initial condition from the beginning.
Thus, in the point of view of the observer, causality is not violated.

The circular logic problem arises from ignorance about the future event horizon.
However, the cosmic horizon has a topological form and cannot be created or removed by any classical means.
The absence or existence of the future event horizon can be determined from the beginning of the Universe.
Thus, the cosmological solutions are divided into two classes: the universe with/without a future event horizon.
How can we find out where we are living in the two classes?
In the previous calculations, the evolution of the Universe was determined from the sum of a second-order differential equation, Eq.~\eqref{HDE:3}, and a boundary condition, Eq.~\eqref{rinfty}.
Given the present data on the Universe, one may trace the Universe backward to identify the initial value of $r_\infty$.
In doing so, future information need not be known.
Even if $r_\infty$ is not known, the future can be predicted from the evolution equation, Eq.~\eqref{HDE:3}, because it is a well-posed differential equation.
Therefore, in principle, we may be able to recognize where we live of the two classes, if the present data on the Universe are known in detail.
The logic so far only holds when the assumptions about the homogeneity of the universe are valid in the Universe because we predict the future evolution on the basis of that logic.
Until we have more accurate data on the present universe, including dark matter and geometry, we cannot tell whether we are living in a universe with a future event horizon or without one.
However, the use of Eq.~\eqref{HDE} for the evolution at the next moment can be justified because the whole evolution of the Universe is governed by a well-posed differential equation, Eq.~\eqref{HDE:3},
and the future evolution of the Universe is only related to the boundary condition $r_\infty$, which can be given at the beginning of the Universe.

\subsection{Action approach to the causality problem}
In the previous section, we dealt with the causality problem by analyzing the equation of motion.
However, if one wants to explain holographic dark energy based on fundamental physics, it should be described by an action, which we do not know as of yet.
In Ref.~\citealp{LiMiao:2012}, by considering the Robertson-Walker metric,
\beqa
\label{Metric:Li}
ds^2= -N^2 (t) dt^2 + a^2 (t) \left[ \frac{dr^2}{1-Kr^2} + r^2 d\Omega^2\right],
\eeqa
Li and Miao proposed an action for the holographic dark energy with the following form:
\beq
\label{S:Li}
S = \frac{1}{16\pi} \int dt \left[ \sqrt{-g} \Big(R - \frac{12d^2}{a^2 L^2} \Big) -\lambda \Big(\dot L + \frac{N}{a} \Big) \right]  + S_M,
\eeq
where $R$ is the Ricci scalar, $\sqrt{-g} = N a^3$, $N^2= -g_{tt}$ is the lapse function of the metric, and  $S_M$ denotes the action of all matter fields.
Note also that $\lambda(t)$ plays the role of a Lagrange multiplier constraining the dynamics of $L(t)$.

With the redefintion of $N dt$ as dt, the equation of motion derived from the action becomes
\begin{equation} \label{EOM:Li}
\frac{\ddot a}{a} = - \frac{\lambda}{6a ^4} - \frac{4\pi}{3} (\rho_M + 3p_M) .
\end{equation}
The auxiliary fields behave as
\beqa
L(t) &=& \int_t^\infty \frac{dt'}{a(t')} + L(\infty), \no
\lambda(t) &=& - \int_0^t dt' \frac{24 a(t') d^2}{L^3(t')} ,
\eeqa
where we assume $\lambda(0)=0$ and $L(\infty) =0$ for the solution in the work.
Note that $\lambda < 0$ for large enough time $(d> 0)$.
When the matter effect is ignorable, Eq.~\eqref{EOM:Li} predicts an accelerating expansion.
Specifically for a holographic-dark-energy-dominated universe with $d<1$, the explicit solution for the Universe predicts a big rip universe:
\begin{equation} \label{BigRip}
a \sim (- t)^{(1-\sqrt{1+ 8d^2})/(3- \sqrt{1+ 8d^2})}  ,
\end{equation}
where the time ends at $t=0$.

The model was analyzed again in the literature~\cite{Li:2012fj} and the results were compared to the observational data in Ref.~\citealp{Cui:2014sma}.
However, the model needs to be revised because it is given as an effective action only on the comoving time $t$, and the metric component $a(t)$ appears explicitly on the action.
To be well-behaving, the action should keep symmetries such as scale invariance and invariance under general coordinate transformations.
The action in Eq.~\eqref{S:Li}, is unsatisfactory in this respect.
An action, including a tachyonic scalar field, was considered in Ref.~\citealp{Rozas-Fernandez:2014vna} to overcome this problem.

\subsection{Other approaches}

Other approaches have tried to avoid the causality problem of holographic dark energy.
Here, we summarize them in their simplest forms.
For example, Nojiri and Odintsov~\cite{Nojiri:2005pu} introduced a $1/R$-like term to the action as an IR behavior at small curvature.
The correction term will be non-negligible when the curvature is small, such as in the present Universe.
In a similar spirit, Gao {\it et al.}~\cite{Gao2009} introduced an energy density proportional to the Ricci scalar curvature, $\rho_\Lambda \propto R$.

An interesting model for the agegraphic dark energy model was also developed by Cai~\cite{Cai2007}.
The model applies the Heisenberg uncertainty principle to the age of the Universe to determine the limit of the distance accuracy, $\delta t = \lambda t_p^{2/3} t^{1/3}$,
where $\lambda $ is a dimensionless constant on the order of unity and $t_p$ is the Planck time.
The quantum uncertainty principle between time and energy present an energy density of metric fluctuations of Minkowski spacetime of the form $\rho_q \propto m_p^2/t^2$.
Choosing the time scale $t$ as the age of the Universe, we may find an energy density, which is similar to that of holographic dark energy.

On the whole, the causality problem of holographic dark energy is resolved at the equation-of-motion level.
However, at the fundamental level of action, the resolution was done only partly, so additional work is necessary.
Therefore, explaining the microscopic origin of dark energy is desirable.
One of the explanations is based on entanglement, which is to be reviewed in Section V.

\section{Entanglement and dark energy}

In this section, we review the idea that holographic dark energy is from a quantum entanglement of the quantum field vacuum of the Universe.
In this model, dark energy is identified as a thermal energy (entanglement energy) associated with
the entanglement entropy $S_{ent}$ of the Universe, and its microscopic origin is suggested.

Quantum entanglement is a kind of  quantum nonlocal correlation that quantum states can have.
A good measure of entanglement for pure states is $S_{ent}$.
Let us consider a bipartite system with subsystems $A$ and $B$, which has a density
matrix $\rho_{AB}$. Then, $S_{ent}$ is given by
 $S_{ent}=-Tr(\rho_A ln \rho_A)$
for a reduced density matrix $\rho_A\equiv Tr_B
\rho_{AB}$.
An interesting fact is that the quantum vacuum of fields in the Universe has natural entanglement due to
the  Reeh-Schlieder theorem~\cite{reeh,werner}.
For example, we consider a Hamiltonian for the massless scalar field $\phi$  in the Minkowski spacetime
 ~\cite{Srednicki1993}
\beq
H=\int d^3 x (|\nabla  \phi(x)|^2+ |\pi(x)|^2),
\eeq
where $\pi(x)$ is the  momentum of the field. We can divide the space into two parts: one inside (A) and the other outside (B) a spherical region like a cosmic horizon. Discretization of a radial coordinate with a UV cut-off gives
an effective Hamiltonian for discretized field oscillators~\cite{Srednicki1993}.
Then, one finds that the leading term in  entanglement entropy between $A$ and $B$ is proportional
to the area of the boundary between the two subsystems.

We expect this area law to hold for a more general space-time.
Therefore, $S_{ent}$ for a spherical region with a radius $r$ can be expressed as
\beq
 \label{Sent}
  S_{ent}=\frac{\beta r^2}{b^2},
  \eeq
where $\beta$ is a constant depending on  the fields, and $b$ is the UV cutoff.
$S_{ent}$ in this form is consistent with the holographic principle.
Numerical calculations gives  a value $\beta\simeq 0.3$ for a massless field ~\cite{Srednicki1993,PhysRevD.52.4512}.
We need to add up the contributions from other $j$ fields with constants $\beta=\beta_j$ and spin degrees of freedom $N_j$ ~\cite{PhysRevD.52.4512}.
Therefore, we expect
\beq
S_{ent}=\sum_j \beta_j N_j\frac{ r^2}{b^2}\equiv \frac{\alpha r^2}{l_P^2},
\eeq
where the Planck length is $l_P=\sqrt{G}$ in $c=\hbar =1$ unit. Choosing the UV cut-off, $b=1/M_P$, gives
\beq
\label{alpha}
\alpha=\frac{1}{8\pi}\sum_j \beta_j N_j.
\eeq
Bianchi showed~\cite{Bianchi:2012br} that a variation of $S_{ent}$ is equal to a variation of the Bekenstein-Hawking entropy
\beq
S_{BH}=\frac{A}{4G}.
\eeq
If we identify $S_{ent}$ with $S_{BH}$, then $\alpha=\pi$.

\begin{figure}[hbtp]
\includegraphics[width=0.32\textwidth]{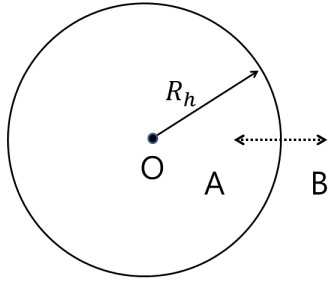}
\caption{ For an observer at $O$ the space inside the cosmic event horizon with radius $R_h$ can be divided into two subspaces, one inside (A) and the other outside (B) the spherical horizon.
The horizon surface $\Sigma$   has the entanglement entropy $S_{ent}\propto {R_h}^2$ and entanglement energy $E_{ent}\equiv  \int_\Sigma T_{ent} dS_{ent}$, which can be dark energy.}
\label{RindlerHorizon}
\end{figure}

What is the relation between $S_{ent}$ and holographic dark energy? In Ref.~\citealp{Lee:2010bg}, the entanglement first law
\beq
\label{firstlaw}
dE_{ent}=k_B T_{ent} dS_{ent}
\eeq
was first derived and was shown to be equivalent to the Einstein equation,
where $dE_{ent}$ is the variation of entanglement energy associated with the local Rindler horizon as shown in Fig.~\ref{RindlerHorizon},
where $k_B$ is the Boltzmann constant and $T_{ent}$ is the horizon temperature.

We assume that the first law holds at a cosmic horizon.
In Ref.~\citealp{LeeLeeKim2007}, we suggested the entanglement energy $E_{ent}$ associated with the future cosmic event horizon is holographic dark energy.
(There are similar suggestions based on the Verlinde's idea ~\cite{Li:2010cj,Zhang:2011,Wei:2011,Easson:2011}.)
Integrating $dE_{ent}$ on the spherical surface ${\Sigma}$ of the event horizon we obtain
 \beq
 \label{eent}
E_{ent}=\int_{\Sigma}  dE_{ent}= k_B T_{ent} \int_{\Sigma}  dS_{ent}= \frac{\alpha R_h}{2 \pi l_P^2},
\eeq
where $T_{ent}= 1 /2\pi k_B R_h $ is the Gibbons-Hawking temperature of the horizon.
The entanglement energy density within $R_h$ is given by
\beq
\label{rho}
\rho_{\Lambda}=\frac{3 E_{ent}}{4 \pi R_h^3}
=\frac{3 \alpha  M_P^2}{\pi R_h^2} ,
\eeq

From the holographic dark energy $\rho_{\Lambda}={3 d^2  M_P^2}/{ R_h^2} $ in Eq.~\eqref{holodark2}, we can obtain  the holographic dark energy parameter
 \beq
\label{d1}
 d=\sqrt{\frac{\alpha}{\pi}}.
 \eeq
That this model makes the holographic dark energy parameter $d$, in principle, calculable, is worth emphasizing.
Before Ref.~\citealp{LeeLeeKim2007}, the parameter $d$ was obtained only by observations.

\begin{figure}[hbtp]
\includegraphics[width=0.32\textwidth]{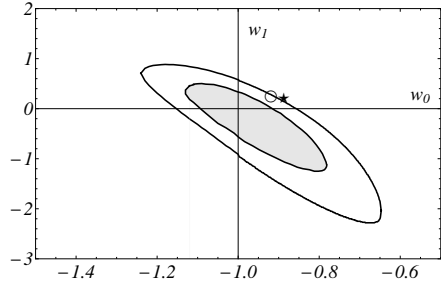}
\caption{ Observational constraints on the dark energy equation of state $w_0$ and $w_1$  from WMAP+BAO+$H_0$+SN. (The data are extracted from
Ref. ~\citealp{Komatsu:2010fb}). The star represents our theoretical prediction with $d=1$ while the circle is for the standard model $(d=0.67)$.}
\label{ObservationalConstraints}
\end{figure}

If we use approximate values for the parameters for field $j$, $\beta_j \simeq 0.3$ and  $\sum_j N_{j}=118$, in the standard model  of particle physics,
we obtain $d=0.67$, which is similar to the observed data $d=0.768^{+0.112}_{-0.068}$~\cite{Wang:2013zca}.
In Fig.~\ref{ObservationalConstraints}, the star and the circle indicate the theoretical predictions with $d=1$ and $d=0.67$, respectively.
Our model provides a way to derive $d$ and $\omega_\Lambda$ in an $ab~initio$ manner for the first time.
However, the above estimate is based on a flat space-time calculation.
In a future work, we need to study quantum entanglement in an expanding universe.

What is the modern version of the entanglement first law?
The reduced density matrix can be represented by a modular Hamiltonian $H_A$
as $\rho_A=e^{-H_A}/Z$, where $Z$ is the partition function.
Then, the variation of entanglement across the boundary is
\beqa
   \delta {{S}_{ent}}&=&-tr(\delta {{\rho }_{A}}\log ({{\rho }_{A}}))-tr({{\rho }_{A}}{{\rho }_{A}}^{-1}\delta {{\rho }_{A}}) \no
  &=&tr(\delta {{\rho }_{A}}{{H}_{A}}) \no
  &=&\delta \left\langle {{H}_{A}} \right\rangle.
\eeqa
Usually, $\rho_A$ is not a thermal state, but in locally flat space-time, $\rho_A$ looks like a thermal state.
We expect this first law to hold for a cosmic horizon with Hawking radiation.

\section{Conclusions}

The problem of identifying the dark energy that explains the cosmic acceleration expansion found in 1998 has become the biggest issue of modern cosmology and theoretical physics. In essence,
the problem of dark energy is a matter of quantum gravity. Therefore, holographic principles as a fundamental principle of quantum gravity will play a very essential role in solving the problem of dark energy.

In this review paper, a holographic dark energy model based on the holographic principle was introduced to analyze the dark energy dynamics.
In this model, the IR cut-off of the UV/IR mixing in the holographic principle suitable for the accelerating expansion of the universe is the size of the horizon of the future event,
which enters into the formula of the holographic dark energy density.
The analysis of the equation of state in this model shows that the case of the constant parameter with $d\lesssim 1$ is favored for the current observational data.

Causality problems and circular logic problems about the holographic dark energy model can be resolved with the action principle and the initial condition of the equations of motion.
Two auxiliary fields, which are introduced from the action principle, make all equations of motion local.
The use of the event horizon as the current cut-off is automatically determined from the equation of motion; therefore, the evolution of the universe is determined by the present initial condition.
However, because the action introduced in Ref.~\citealp{LiMiao:2012} does not have the covariant form with full symmetry, it still remains unsatisfactory.

Jacobson's idea linking gravity with thermodynamics requires a microscopic degree of freedom.
Quantum entanglement of the quantum field vacuum in curved space-time is a natural candidate for the microscopic degrees of freedom for the thermodynamic version of gravity.
If we extend this idea to the dark energy problem, we can easily arrive at the holographic dark energy model associated with an entanglement of cosmic horizons.
Interestingly, a rough estimate in this model gives a holographic dark energy comparable to the current observational data.
Additionally, that the holographic dark energy model associated with an entanglement of cosmic horizons provides a theoretical basis for the calculation of dark energy parameter $d$ is worth emphasizing.
However, to get the exact equation of state we still need more precise calculations for the entanglement in curved space-time.

\section*{ACKNOWLEDGEMENT}
This work was supported by the Daejin University Research Grants in 2018.

\end{document}